\documentclass[twocolumn,showpacs,preprintnumbers,amsmath,amssymb]{revtex4}
\usepackage{lipsum}
\usepackage{tabularx}
\usepackage{array}
\usepackage{mathrsfs}
\usepackage{amssymb}
\usepackage{amsmath}
\usepackage{cases}  
\usepackage{graphicx}
\usepackage{subfigure} 
\usepackage{dcolumn}
\usepackage{bm}
\usepackage{verbatim} 
\usepackage[dvipdfm,pdfstartview=FitH,colorlinks,linkcolor=blue,anchorcolor=blue,citecolor=blue]{hyperref}
\usepackage{amsthm}  

\begin{document}

\title{Quantum key distribution with on-chip dissipative Kerr soliton}

\author{Fang-Xiang Wang$^{1,3,5}$, Weiqiang Wang$^{2,4,*}$, Rui Niu$^{1,3}$, Xinyu Wang$^{2,4}$, Chang-Ling Zou$^{1,3}$, Chun-Hua Dong$^{1,3,*}$, Brent E. Little$^{2}$, Sai T. Chu$^{6}$, Hang Liu$^{1,3,5}$, Penglei Hao$^{7}$, Shufeng Liu$^{7}$, Shuang Wang$^{1,3,5}$, Zhen-Qiang Yin$^{1,3,5}$, De-Yong He$^{1,3,5}$, Wenfu Zhang$^{2,4,*}$, Wei Zhao$^{2,4}$, Zheng-Fu Han$^{1,3,5}$, Guang-Can Guo$^{1,3,5}$, and Wei Chen$^{1,3,5,*}$}

\affiliation{$^1$CAS Key Lab of Quantum Information, University of Science and Technology of China, Hefei 230026, China\\
	$^2$State Key Laboratory of Transient Optics and Photonics, Xi'an Institute of Optics and Precision Mechanics (XIOPM), Chinese Academy of Sciences, Xi'an 710119, China\\
	$^3$CAS Center For Excellence in Quantum Information and Quantum Physics, University of Science and Technology of China, Hefei, Anhui 230026, China\\
	$^4$University of Chinese Academy of Sciences, Beijing 100049, China\\
	$^5$State Key Laboratory of Cryptology, P. O. Box 5159, Beijing 100878, China\\
	$^6$Department of Physics and Materials Science, City University of Hong Kong, Hong Kong, China\\
	$^7$Anhui Qasky Quantum Technology Co., Ltd., Wuhu 241002, China}


\begin{abstract}
Quantum key distribution (QKD) can distribute symmetric key bits between remote legitimate users with the guarantee of quantum mechanics principles. For practical applications, there are increasing attentions to integrating the source, detectors, and modulators on a photonic chip. Here we introduce the Kerr dissipative soliton in a microresonator as the light source for GHz QKD systems, and the proof-of-principle experiment demonstrates the parallel QKD can be achieved using the coherent comb lines from the soliton. We also verify the performance of parallel QKD by using the off-the-shelf wavelength division modules and the on-chip soliton source. Since the on-chip soliton can provide hundreds of carriers covering C and L bands, the results exhibit the feasibility to achieve ultra-high rate secure key bits through massively parallel QKD, especially incorporating with the thriving photonic integrated circuit technology.
\end{abstract}


\maketitle

\section{Introduction}
\label{introduction}

Quantum key distribution (QKD) can distribute symmetric key bits between remote legitimate users \cite{Bennett2014}. Benefiting from its quantum generating process, QKD can achieve information-theoretical security, which is not based on the computation assumption \cite{Scarani2009,Lo2014}. A lot of research efforts have been devoted to getting higher secure key rate (SKR), longer secure transmission distance in a QKD point-to-point link \cite{Wangs2012,Korzh2015,Yuan2018,Boaron2018a,Boaron2018b}, as well as more flexible connections and higher user capacities in QKD networks \cite{Chen2009,Wangs2010,Sasaki2011,Wangs2014}. The photonic layer, which can generate, modulate and detect photon pulses at the single photon level, is the kernel of QKD. Although the clocking rate of practical QKD systems have exceeded 1 GHz and the SKR can be more than 1 Mbps at the fiber quantum channel of 50 km \cite{Tanaka2012}, to boost the SKR is still the most pressing task in QKD research. Wavelength division multiplexing (WDM) is an essential technique in increasing the SKR in an end-to-end QKD system \cite{Yoshino2012,Yoshino2013,Patel2014}, and routing quantum signals in QKD networks \cite{Wangs2010,Sasaki2011}. The emerging achievements of integrated photonic QKD chips \cite{Sibson2017,Price2018,Bunandar2018}, and single photon detector arrays \cite{Miyajima2018} have exhibited the potential to integrate high speed and large-scale QKD transceivers on a single chip.

The maturest light source used in current real-life QKD systems is the weak coherent source (WCS), and arrays of lasers are deployed in the scenario of using WDM. However, for integrated QKD transceiver chips \cite{Sibson2017,Bunandar2018} with massive parallel and high-speed code units, the locked laser sources are necessary but very challenging. Even for traditional off-chip QKD systems, such a light source is cumbersome or costs too much. Recently, there is emerging a new research field on the optical frequency comb generation in microcavities \cite{DelHaye2007,Kippenberg2011,Weiner2017,Kippenberg2018}, which provides abundant individual frequency resources, simultaneously, in a wide bandwidth with only a single external laser pump. The dissipative Kerr soliton (DKS) soliton source \cite{Herr2013,Li2017,Kippenberg2018} can generate high signal-to-noise ratio (SNR) individual optical comb lines in a wide bandwidth ($\sim 150\,\mathrm{nm}$) covering C and L bands, with the frequency spacing of about $50\,\mathrm{GHz}$ or $100\,\mathrm{GHz}$, which is suitable for wavelength multiplexing in commercial optical communication systems. Such an attractive source has already been applied in distance measurement \cite{Suh2018,Trocha2018} and classical optical communication, which indicates the potential capacity of transmitting tens of Tbps bitstreams using the compact device to replace the laser arrays \cite{Marin-Palomo2017}. 

In this work, we demonstrate the feasibility of implementing ultra-high-key-rate QKD system using an on-chip DKS source and WDM in two scenarios. In the first scenario, a set of end-to-end QKD system with the light pulse repetition rate of $1\,$GHz runs using one frequency channel of the comb lines. In the second scenario, two sets of independent $1\,$GHz QKD systems run simultaneously within different communication channels. According to these two scenarios, we realized an SKR of more than $200\,$kbps with a set of QKD system at the distance of $25\,$km. The massive narrow line width and frequency locked comb source provided by on-chip soliton has exploited the prospect of implementing chip-scale QKD transceivers with the gigabit-per-second SKR by combing with the integrated photonics circuits and the advanced multiplexing technologies \cite{Stern2018}.

\section{Concept and dissipative Kerr soliton}
\label{concept}

\begin{figure*}
	\centering \resizebox{16cm}{6.37cm}{\includegraphics{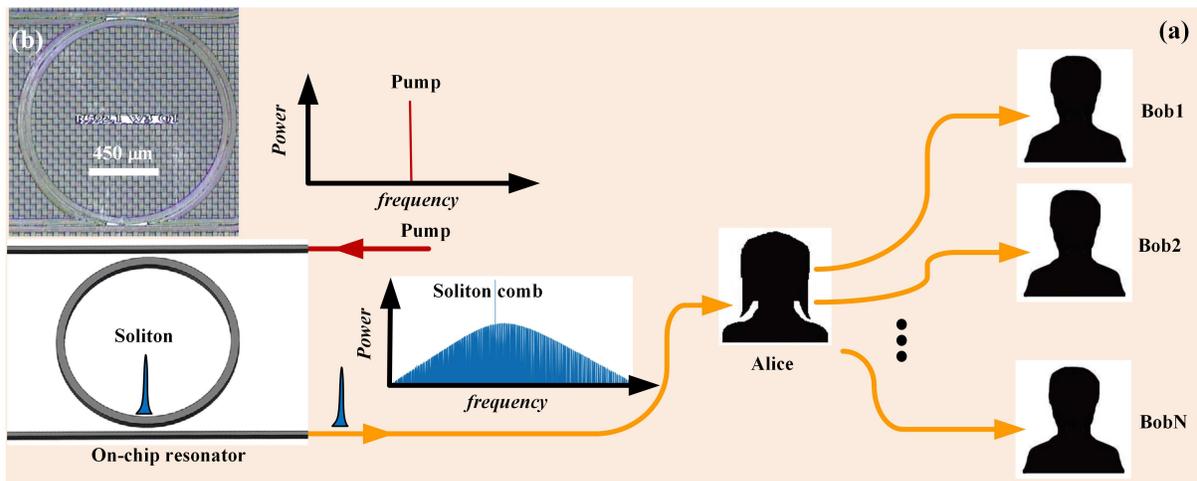}}
	\caption{(a) The conceptual illustration of the soliton based QKD. An external narrow-linewidth pump laser generates the dissipative Kerr soliton within the on-chip resonator. The soliton comb is then be used as the light source to realize the ultra-high speed (multi-user) QKD system by frequency multiplexing/demultiplexing. (b) The micrograph of the microresonator.}
	\label{fig:1} 
\end{figure*}

\noindent\textbf{Concept.} 
Figure$\,$\ref{fig:1}(a) schematically illustrates the concept of QKD networks based on the on-chip DKS source. In an integrated microring resonator, there are whispering gallery modes with almost equal frequency spacing and very high-quality factor ($Q$) \cite{Vahala2004}. Due to the competition between the external pump laser, cavity dissipation and the intracavity Kerr effect, the resonator could evolve into a stable soliton state that a pulse circulating in the resonator with fixed pulse shape \cite{Herr2013,Li2017,Kippenberg2018}. Through the external bus waveguide, the soliton could be coupled out and serves as the laser sources for the QKD experiments. In the frequency domain, the stable pulse corresponds to a comb, consisting of many coherent laser lines with equal frequency spacing. Therefore, ultra-high speed and multi-user  QKD network could potentially be realized through wavelength multiplexing. 

\noindent\textbf{Dissipative Kerr soliton.} 
Our work uses an integrated high-index doped silica glass microresonator \cite{Wang2018a,Wang2018b} for DKS frequency comb generation, as shown in Fig. \ref{fig:1}(b). By carefully designing the waveguide cross-section of the microring resonator, a flat anomalous group velocity dispersion in communication bands is realized, which is beneficial for broadband frequency comb generation. By gluing the photonic integrated chip with optical fiber, the microring resonator can be efficiently excited and the DKS can be collected directly through the fiber. Additionally, our device is packaged with a thermoelectric cooler (TEC) for enhancing the environmental adaptability of the DKS source \cite{Wang2018b}.

\begin{figure*}
	\centering \resizebox{17cm}{11.61cm}{\includegraphics{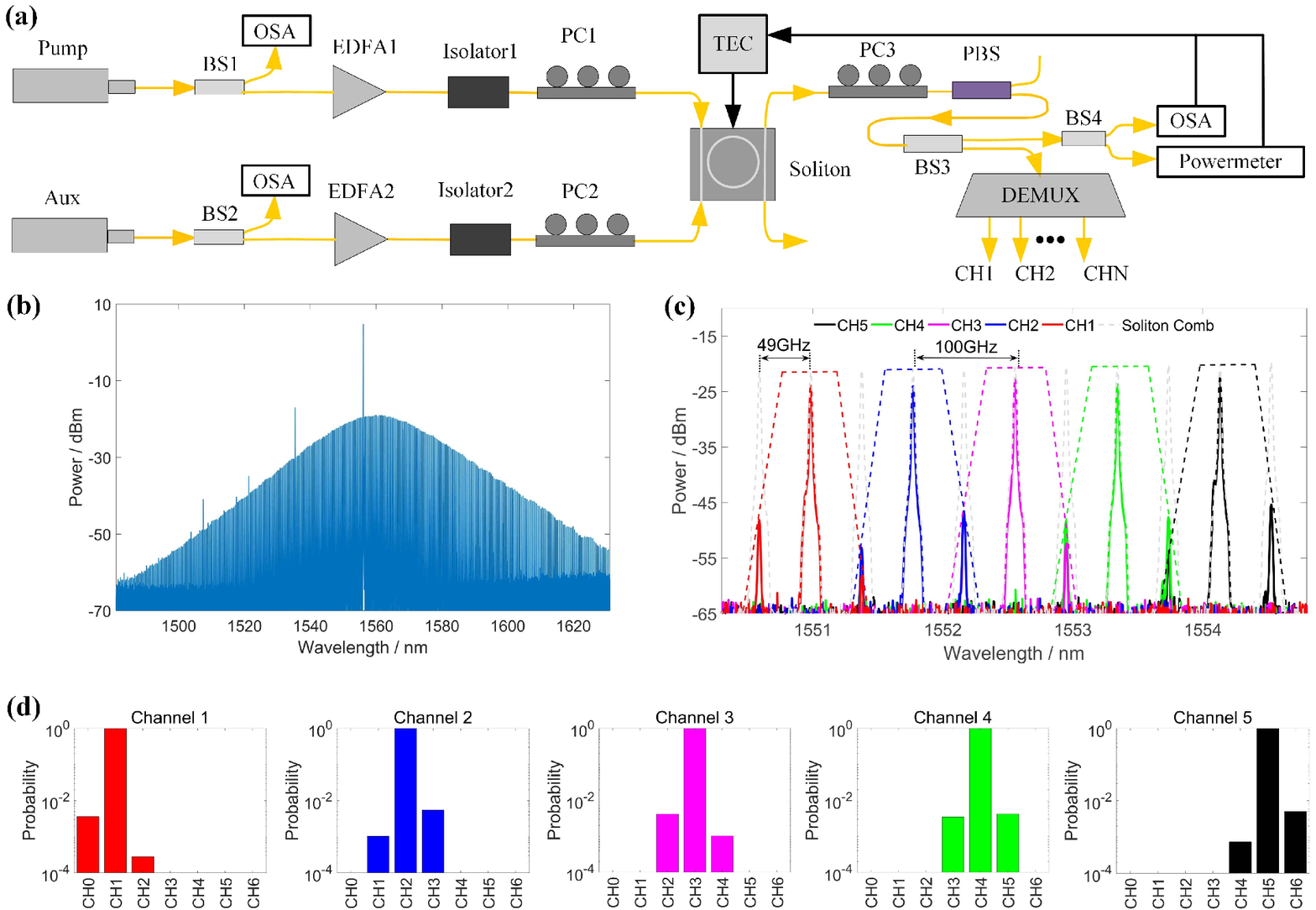}}
	\caption{The dissipative Kerr soliton generation and evaluation. (a) The experimental setup of the dissipative Kerr soliton source for frequency multiplexing and demultiplexing. The pump (Pump) and auxiliary (Aux) lasers are used to generate the DKS and to timely balance the heat fluctuation within the microresonator, respectively. The optical spectrum analyzer (OSA) and power meter are used to monitor the output spectrum and power of the microresonator, respectively. The temperature of the microresonator is manually tuned by a thermoelectric cooler (TEC) with the feedback of the OSA and power meter. The laser polarizations before and after the microresonator are controlled by polarization controllers (PC1-PC3). The dissipative Kerr soliton is then be demultiplexed. (b) Typical measured power spectra of the soliton comb. (c) The measured power spectra of the output channels of the demultiplexing (DEMUX) in (a). (d) Channel crosstalk of the DEMUX. For left to right: the power spectra crosstalk of CH1 to CH5. }
	\label{fig:2} 
\end{figure*}

The experimental setup of our integrated DKS source is shown in Fig.$\,$\ref{fig:2}(a). The radius of the microring is $592.1\,\mathrm{\mu m}$, corresponding to a free spectral range of about $49\,\mathrm{GHz}$, and the typical Q-factor of our device is about $1.7\times 10^6$. The waveguide cross-section of the microring resonator is carefully designed as $2\,\mathrm{\mu m}\times 3\,\mathrm{\mu m}$. An external pump laser is amplified by an erbium-doped fiber amplifier (EDFA) and coupled to the microring resonator. Due to the ultra-high Q-factor enhanced Kerr nonlinearity in the microring, the whispering gallery modes start to lase through the hyperparametric oscillation \cite{DelHaye2007}. By carefully tunning the relative frequency detuning of the pump laser and the microring resonance, the DKS could be realized \cite{Herr2013,Herr2015,Kippenberg2018}. Here, we adopt the auxiliary laser assisted soliton switching method to kick the microring to the multi-soliton state \cite{Lu2018,Niu2018}, where the auxiliary laser is used to balance the intra-cavity heat. The frequency of the pump laser is fixed at 1556.09nm and the intensity output from EDFA is 3.5W. The intensity of the auxiliary laser is 3.2W. Since the comb spacing of the N-soliton state is random and uncertain, we manually switch the system to the single soliton state for a full comb by annihilating the number of soliton step by step \cite{Guo2016,Lu2018}. It is worth noting that, the auxiliary laser is coupled into the microring counter-propagate with respect to the soliton. The polarization of the laser is also orthogonal to that of the soliton. Thus, the auxiliary laser can be filtrated by a polarizing beam splitter (PBS) and has little influence on the soliton source. The soliton comb states are stable over several hours by maintaining the comb power with frequency control of the auxiliary laser or with temperature control of the TEC. The temperature resolution and stability of the TEC are 0.01 degree/1 $\Omega$ and 0.002 degree, respectively.

Figure$\,$\ref{fig:2}(b) presents the power spectra of the single-soliton state achieved in our experiment. The soliton exhibits a 3-dB spectral bandwidth as broad as $3.2\,\mathrm{THz}$ and more than 200 comb lines with $20\,$dB SNR. The single-soliton state shows features, i.e. the exactly equal comb-spacing and the smooth spectrum envelop, that are attractive for ultra-high speed optical communication with frequency multiplexing \cite{Marin-Palomo2017}.

\section{Soliton-based wavelength multiplexing QKD}
\label{QKD}

To demonstrate the feasibility of the integrated DKS source for the QKD applications, the wavelength multiplexing of the DKS was firstly tested. The power spectra crosstalks of five channels (CH1-CH5) of the demultiplexing (DEMUX) are shown in Fig.$\,$\ref{fig:2}(c). The soliton comb line interval is $49\,\mathrm{GHz}$ and the channel interval is $100\,\mathrm{GHz}$. The power spectra crosstalks of the DEMUX are shown in Fig.$\,$\ref{fig:2}(d) (from left to right: CH1 to CH5). Our experimental results indicate that the cross talk between adjacent and non-adjacent channels are less than $20\,$dB and $40\,$dB, respectively, for CH1-CH5. Therefore, the DKS  promises a high extinction ratio (ER) comb source for wavelength multiplexing QKD.

\begin{figure*}
	\centering \resizebox{16cm}{6.03cm}{\includegraphics{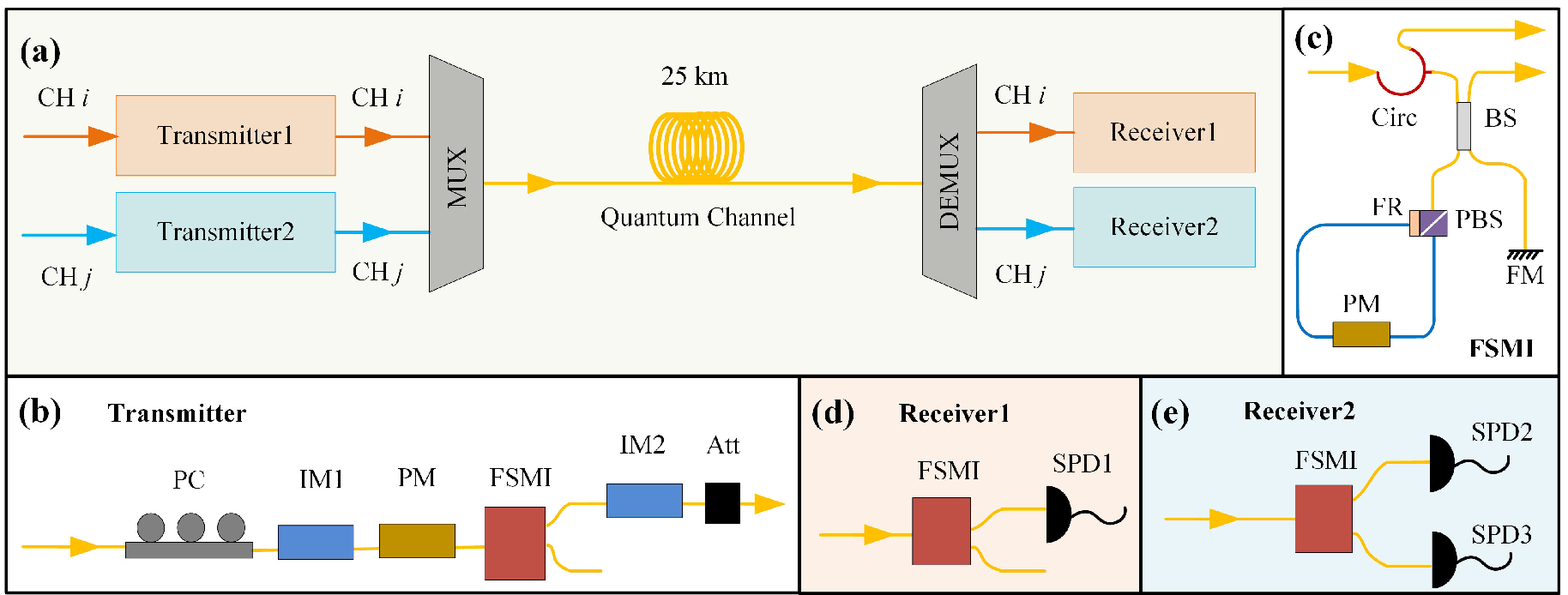}}
	\caption{The experimental setup of the soliton-based QKD system. (a) The schematic setup of the multi-channel QKD system. Two QKD systems run independently with different channels. All channels are multiplexed and demultiplexed by the dense wavelength division multiplexing (DWDM). The length of the quantum channel is 25 km. (b) The encoding setup of the transmitters. (c) The structure of the Faraday-Sagnac-Michelson interferometer (FSMI). The blue line of the Sagnac loop is the polarization maintaining fiber. (d) and (e) The decoding setup of receivers with single and double single-photon detectors (SPDs), respectively. (DE)MUX: (de)multiplexing; BS: non-polarizing beam splitter; PBS: polarizing beam splitter; FR: Faraday rotator; FM: Faraday mirror; PM: phase modulator; IM: intensity modulator; Att: attenuator; Cir: circulator. }
	\label{fig:3} 
\end{figure*}

QKD systems running in WDM scenarios are then evaluated. The experimental setup for the proof-of-principle demonstration of the DKS-based QKD is shown in Fig.$\,$\ref{fig:3}(a). For each transmitter on Alice's side, a DEMUX is connected to the DKS source to select a working bandwidth, as shown in Fig.$\,$\ref{fig:2}(a). The selected comb lines output from different channels of the DEMUX are actually continuous lasers. Hence, an intensity modulator (IM) (IM1 in Fig.$\,$\ref{fig:3}(b)) is used to chop the light into pulses. The polarization controller (PC) connected before IM1 is used to get the maximum ER of light pulses. A phase modulator (PM) is cascaded to randomize the global phase of each pulse \cite{Cao2015}. This randomization is not only the requirement of the weak coherent light itself, but also to eliminate the phase correlations between the different components of the combs. Then the classical random bits are phase-encoded onto the photons based on BB84 protocol with decoy-state method \cite{Wangxb2005,Lo2005}.

The encoding processes are implemented by a Faraday-Sagnac-Michelson interferometer (FSMI), in which the Faraday mirror (FM) in its long arm is replaced by a Sagnac configuration with a PM inserted \cite{Wangs2018}. The quantum states carried by the photons can be divided into two mutually unbiased bases (MUB) X and Y, which are defined as $(1/\sqrt{2})(|s\rangle\pm|l\rangle)$ and $(1/\sqrt{2})(|s\rangle\pm i|l\rangle)$, respectively. $|s\rangle$ and $|l\rangle$ here are the pulses passing through the short and long paths of the FSMI, respectively, and the phase difference between the two pulses are randomly modulated by the PM in the Sagnac configuration with the phases of $\{0,\pi/2,\pi,3\pi/2\}$. The pulses exporting from FSMI are randomly modulated to different intensities by IM2 to meet the requirement of decoy state method. The pulses are then attenuated to the single-photon level by an attenuator (ATT) before entering the quantum channel.

After transmitting through a quantum channel of 25km fiber, another DEMUX is used to demultiplex signals from different channels into corresponding QKD receivers in Bob. The structures of the two QKD receivers (Receiver1 and Receiver2) are shown in Figs.$\,$\ref{fig:3}(d) and \ref{fig:3}(e), respectively. The decoders of the QKD receivers are FSMI units which have the same structure as that in the transmitters in Alice. The receiver measures the qubits either in the $X$ basis or $Y$ basis by randomly modulating the PM in its decoder with the probabilities $p_{X}$ and $p_{Y}$, respectively. The decoding results are detected by single-photon detectors (SPD). The only difference between the two receivers is the number of SPD. It is worth noting that the single-SPD scheme has the same security with the double-SPD scheme since the decoders use four-phase active modulation. Since the SPD is one of the most expensive components in QKD system, the users can make flexible selections between 3dB detection efficiency sacrifice and the cost reduction, which may be valuable in QKD networks with massive users.

\section{Performance}
\label{performance}

Our DKS-based QKD systems are implemented in two different scenarios. In the first scenario, the performances of single-channel QKD systems are characterized. Thus, only one channel of the DKS spectrum in Fig.$\,$\ref{fig:2}(a) is used to implement the QKD. In the second scenario, multi-channel QKD systems are demonstrated by employing two communication channels of the DKS. In this scenario, two independent QKD systems are running simultaneously to estimate the crosstalk between different channels.

The QKD system utilizes one signal and two decoy states and the clock rates of the QKD system is 1GHz. The average photon number per transmitted pulse of the first QKD system TR1 (Transmitter1\&Receiver1) is $\lambda_{\mu,1}=0.50$, $\lambda_{\nu_{1},1}=0.16$ and $\lambda_{\nu_{2},1}=0.008$, respectively, where $\mu$ represents the signal state, and $\nu_{1}$ and $\nu_{2}$ represent the two decoy states. The corresponding values of the second system TR2 (Transmitter2\&Receiver2) are $\lambda_{\mu,2}=0.60$, $\lambda_{\nu_{1},2}=0.06$ and $\lambda_{\nu_{2},2}=0.008$, respectively. The sending ratio of $\mu$, $\nu_{1}$ and $\nu_{2}$ is 29:2:1 for both TR1 and TR2. The probabilities of choosing $X$ basis or $Y$ basis are $p_{X}=p_{Y}=1/2$. The detection efficiencies of the SPDs of TR1 and TR2 are larger than 15\%. The fiber loss of the quantum channel is 0.2 dB/km. The total insert loss of the MUX and DEMUX between the Transmitter and Receiver in Fig. \ref{fig:3}(a) depends on the specific channel and varies from 4.7 dB to 5.2 dB. The post-processing efficiency of TR2 is higher than that of TR1 as there are two buffers in TR2 while there is only one buffer in TR1. The Transmitter and Receiver are synchronized by a reference laser through an equal-length (25 km) reference fiber channel (not shown in Fig. $\,$\ref{fig:3}). The wavelength of the reference laser differs from that of the quantum channel.

\begin{figure*}
	\centering \resizebox{15cm}{7.58cm}{\includegraphics{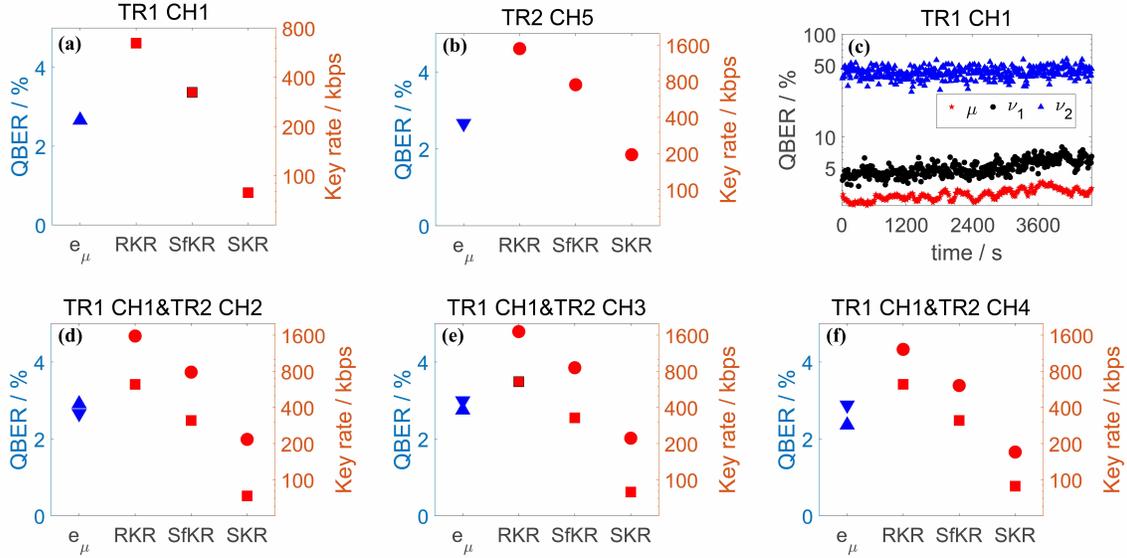}}
	\caption{The QBER of the signal state ($e_{\mu}$), raw key rate (RKR), sifted key rate (SfKR) and secret key rate (SKR) of the (a)-(b) single-channel and (d)-(f) multi-channel QKD systems. The regular triangle and square represent the experimental data of TR1. The inverted triangle and circle represent the experimental data of TR2. (c) The one-hour real-time QBER of TR1 by CH1. The red star, black circle, and blue regular triangle represent $\mu$, $\nu_{1}$ and $\nu_{2}$, respectively. The sending ratio of $\mu$, $\nu_{1}$ and $\nu_{2}$ of the transmitter is 29:2:1. TR1 CH1\&TR2 CH2 means that system TR1 runs within CH1, while TR2 runs within CH2 simultaneously but independently.}
	\label{fig:4} 
\end{figure*}

In the first scenario, the single-channel QKD systems are verified. We run TR1 and TR2 independently within a single channel of the DEMUX in Fig.$\,$\ref{fig:2}(a). The quantum bit error rates (QBERs) of TR1 and TR2 are $2.66\pm 0.48$ and $2.66\pm 0.33$, respectively, within one hour, as shown in Figs.$\,$\ref{fig:4}(a) and \ref{fig:4}(b). The average SKRs of TR1 and TR2 are 78.764kbps and 195.360kbps, respectively, where all time overhead including the post-processing procedure has been considered. The post-processing efficiencies of TR2 is higher than that of TR1, thus the SKR of TR2 is larger than twice of that of TR1. The one-hour real-time QBER of TR1 for signal and decoy states are shown in Fig.$\,$\ref{fig:4}(c). It shows that QBERs of all states increase little after one hour's running and demonstrates that the system is stable for the single-channel scenario.

In the second scenario, we verify the multi-channel QKD systems. Two QKD systems (TR1 and TR2) are running simultaneously but independently within different channels. The experimental results show that there is no significant difference for QBER and SKR between adjacent and nonadjacent channels, as shown in Figs.$\,$\ref{fig:4}(d)-\ref{fig:4}(f). It shows that the average QBERs of the signal state are within $3\%$ and the SKRs are in the same level for all multi-channel verification experiments. There is no significant difference between the single and multi-channel experiments and it demonstrates that the performances of the first and second scenarios are well-matched.

\section{Discussion}
\label{discussion}

\noindent Increasing the SKR is a demanding goal of the QKD community. Frequency multiplexing is one of the most common approaches. Therefore, the development of integrated optical chips not only makes on-chip QKD becoming a prospective branch to realize miniaturized and affordable QKD systems \cite{Sibson2017,Price2018,Bunandar2018}, but offers a new QKD platform with dense multiplexing that is much easier to implement and costs much less. The microresonator soliton source is part of an integrated optical chip and supports dense multiplexing in the domain of frequency. More importantly, the soliton is much superior than multi-laser sources. For a multi-laser source, each laser should be individually monitored to fit the corresponding communication channel. However, for DKS, as $f_N=f_0+N\Delta f$, it only needs to monitor the pump laser and comb spacing to make sure hundreds of frequency comb lines fitting the corresponding communication channels, where $f_0$ is the frequency of the pump laser and $\Delta f$ is the comb frequency spacing. Based on this work, it becomes possible to achieve Gbps SKR QKD system with integrated microresonator soliton source by combining with the ultra-fast on-chip modulation technologies\cite{Stern2018}.

It is worth emphasizing that the quantum frequency combs have achieved rapid development in recent years, which may bring the powerful approach for efficient generation and flexible manipulation of photon resource in quantum information research \cite{Kues2019}. It is certain that the most valuable feature of the quantum frequency comb is the temporal and frequency correlations of the photons generated in the same spatial mode. However, the correlation is not necessary for QKD systems with the WCS, in which the frequency combs of the photons are used as individual carriers of quantum states. 

Although our work has not utilized the intrinsic coherent feature of the combs, it is the first step to combine this powerful resource into the real-life sessions like quantum communication. We have demonstrated the QKD system using the on-chip dissipative Kerr soliton sources instead of the off-the-shelf distributed feedback (DFB) lasers. Although, the performance of the QKD system in our work is still not amazing comparing with the current QKD system and the Watt-level pump light for generating on-chip DKS in our work limits its applications outside of the lab, the convenience and stability can be expected when using massively parallel QKD system. And fortunately, the pump power required can be effectively reduced by choosing different materials and improving the Q-factor of the on-chip micro-cavity. By utilizing hybrid integration technologies, the laser source and the micro-cavity have been integrated onto a single chip to generate the on-chip soliton with the electrical driving power less than 100 $mW$, as well as the good stability in the optical spectrum and power \cite{Stern2018}. The rapid developments have revealed the prospects of the on-chip soliton sources and its potential to be integrated into practical applications like QKD.

\section{Conclusion}
In conclusion, We have demonstrated the feasibility of realizing an ultra-high SKR QKD system based on the on-chip DKS source. Through the wavelength multiplexing, the single-channel and multi-channel QKD experiments are performed, and the experiments demonstrated a SKR about $200\,\mathrm{kbps}$ by one channel. Our experiments reveal the potential of DKS-based QKD system for high SKR based on the massive and parallel communication channels. Combining with the exciting progresses achieved in the quantum photonic integrated circuits, the DKS source, high-speed modulators, and high-efficiency SPDs could be potentially realized on a single chip. Such a QKD chip indicates the potential of Gbps SKR and possesses the advantages of compact, low-cost, robust and high efficiency, and will greatly promote the popularization of QKD in personal equipment and networks.

\section*{Acknowledgment}
\label{acknowledgment}
	This work has been supported by the National Key Research And Development Program of China (Grant No. 2018YFA0306400); the National Natural Science Foundation of China (Grant Nos. 61627820, 61475188, 61622506, 61635013, 61675189); the Strategic Priority Research Program of the Chinese Academy of Sciences (Grant No. XDB24030600); and the Anhui Initiative in Quantum Information Technologies.


\begin{thebibliography}{00}
	\bibitem{Bennett2014} C. H. Bennett and G. Brassard, Theor. Comput. Sci. \textbf{560}, 7--11 (2014).
	
	\bibitem{Scarani2009} V. Scarani, H. Bechmann-Pasquinucci, N. J. Cerf, M. Dusek, N. Lütkenhaus, and M. Peev, Rev. Mod. Phys. \textbf{81}, 1301--1350 (2009).
	
	\bibitem{Lo2014} H.-K. Lo, M. Curty, and K. Tamaki, Nat. Photonics \textbf{8}, 595--604 (2014).
	
	\bibitem{Wangs2012} S. Wang, W. Chen, J.-F. Guo, Z.-Q. Yin, H.-W. Li, Z. Zhou, G.-C. Guo, and Z.-F. Han, Opt. Lett. \textbf{37}, 1008--1010 (2012).
	
	\bibitem{Korzh2015} B. Korzh, C. C. W. Lim, R. Houlmann, N. Gisin, M. J. Li, D. Nolan, B. Sanguinetti, R. Thew, and H. Zbinden, Nat. Photonics \textbf{9}, 163--168 (2015).
	
	\bibitem{Yuan2018} Z. Yuan, A. Plews, R. Takahashi, K. Doi, W. Tam, A. W. Sharpe, A. R. Dixon, E. Lavelle, J. F. Dynes, A. Murakami, M. Kujiraoka, M. Lucamarini, Y. Tanizawa, H. Sato, and A. J. Shields, J. Light. Technol. \textbf{36}, 3427--3433 (2018).
	
	\bibitem{Boaron2018a} A. Boaron, G. Boso, D. Rusca, C. Vulliez, C. Autebert, M. Caloz, M. Perrenoud, G. Gras, F. Bussieres, M.-J. Li, D. Nolan, A. Martin, and H. Zbinden, Phys. Rev. Lett. \textbf{121}, 190502 (2018).
	
	\bibitem{Boaron2018b} A. Boaron, B. Korzh, R. Houlmann, G. Boso, D. Rusca, S. Gray, M. J. Li, D. Nolan, A. Martin, and H. Zbinden, Appl. Phys. Lett. \textbf{112}, 1--5 (2018).
	
	\bibitem{Chen2009} Wei Chen, Zheng-Fu Han, Tao Zhang, Hao Wen, Zhen-Qiang Yin, Fang-Xing Xu, Qing-Lin Wu, Yun Liu, Yang Zhang, Xiao-Fan Mo, You-Zhen Gui, Guo Wei, and Guang-Can Guo, IEEE Photonics Technol. Lett. \textbf{21}, 575--577 (2009).
	
	\bibitem{Wangs2010} S. Wang, W. Chen, Z.-Q. Yin, Y. Zhang, T. Zhang, H.-W. Li, F.-X. Xu, Z. Zhou, Y. Yang, D.-J. Huang, L.-J. Zhang, F.-Y. Li, D. Liu, Y.-G. Wang, G.-C. Guo, and Z.-F. Han, Opt. Lett. \textbf{35}, 2454--2456 (2010).
	
	\bibitem{Sasaki2011} M. Sasaki, M. Fujiwara, H. Ishizuka, W. Klaus, K. Wakui, M. Takeoka, S. Miki, T. Yamashita, Z. Wang, A. Tanaka, K. Yoshino, Y. Nambu, S. Takahashi, A. Tajima, A. Tomita, T. Domeki, T. Hasegawa, Y. Sakai, H. Kobayashi, T. Asai, K. Shimizu, T. Tokura, T. Tsurumaru, M. Matsui, T. Honjo, K. Tamaki, H. Takesue, Y. Tokura, J. F. Dynes, A. R. Dixon, A. W. Sharpe, Z. L. Yuan, A. J. Shields, S. Uchikoga, M. Legré, S. Robyr, P. Trinkler, L. Monat, J.-B. Page, G. Ribordy, A. Poppe, A. Allacher, O. Maurhart, T. Langer, M. Peev, and A. Zeilinger, Opt. Express \textbf{19}, 10387--10409 (2011).
	
	\bibitem{Wangs2014} S. Wang, W. Chen, Z.-Q. Yin, H.-W. Li, D.-Y. He, Y.-H. Li, Z. Zhou, X.-T. Song, F.-Y. Li, D. Wang, H. Chen, Y.-G. Han, J.-Z. Huang, J.-F. Guo, P.-L. Hao, M. Li, C.-M. Zhang, D. Liu, W.-Y. Liang, C.-H. Miao, P. Wu, G.-C. Guo, and Z.-F. Han, Opt. Express \textbf{22}, 21739--21756 (2014).
	
	\bibitem{Tanaka2012} A. Tanaka, M. Fujiwara, K. Yoshino, S. Takahashi, Y. Nambu, A. Tomita, S. Miki, T. Yamashita, Z. Wang, M. Sasaki, and A. Tajima, IEEE J. Quantum Electron. \textbf{48}, 542--550 (2012).
	
	\bibitem{Yoshino2012} K. Yoshino, M. Fujiwara, A. Tanaka, S. Takahashi, Y. Nambu, A. Tomita, S. Miki, T. Yamashita, Z. Wang, M. Sasaki, and A. Tajima, Opt. Lett. \textbf{37}, 223--225 (2012).
	
	\bibitem{Yoshino2013} K. Yoshino, T. Ochi, M. Fujiwara, M. Sasaki, and A. Tajima, Opt. Express \textbf{21}, 31395--31401 (2013).
	
	\bibitem{Patel2014} K. A. Patel, J. F. Dynes, M. Lucamarini, I. Choi, A. W. Sharpe, Z. L. Yuan, R. V. Penty, and A. J. Shields, Appl. Phys. Lett. \textbf{104}, 051123 (2014).
	
	\bibitem{Sibson2017} P. Sibson, C. Erven, M. Godfrey, S. Miki, T. Yamashita, M. Fujiwara, M. Sasaki, H. Terai, M. G. Tanner, C. M. Natarajan, R. H. Hadfield, J. L. O’Brien, and M. G. Thompson, Nat. Commun. \textbf{8}, 13984 (2017).
	
	\bibitem{Price2018} A. B. Price, P. Sibson, C. Erven, J. G. Rarity, and M. G. Thompson, in Conference on Lasers and Electro-Optics (OSA, 2018), p. JTh2A.24. DOI: 10.1364/CLEO\_AT.2018.JTh2A.24
	
	\bibitem{Bunandar2018} D. Bunandar, A. Lentine, C. Lee, H. Cai, C. M. Long, N. Boynton, N. Martinez, C. DeRose, C. Chen, M. Grein, D. Trotter, A. Starbuck, A. Pomerene, S. Hamilton, F. N. C. Wong, R. Camacho, P. Davids, J. Urayama, and D. Englund, Phys. Rev. X \textbf{8}, 021009 (2018).
	
	\bibitem{Miyajima2018} S. Miyajima, M. Yabuno, S. Miki, T. Yamashita, and H. Terai, Opt. Express \textbf{26}, 29045--29054 (2018).
	
	\bibitem{DelHaye2007} P. Del'Haye, A. Schliesser, O. Arcizet, T. Wilken, R. Holzwarth, and T. J. Kippenberg, Nature \textbf{450}, 1214--1217 (2007).
	
	\bibitem{Kippenberg2011} T. J. Kippenberg, R. Holzwarth, and S. A. Diddams, Science \textbf{332}, 555--559 (2011).
	
	\bibitem{Weiner2017} A. M. Weiner, Nat. Photonics \textbf{11}, 533--535 (2017).
	
	\bibitem{Kippenberg2018} T. J. Kippenberg, A. L. Gaeta, M. Lipson, and M. L. Gorodetsky, Science \textbf{361}, eaan8083 (2018).
	
	\bibitem{Herr2013} T. Herr, V. Brasch, J. D. Jost, C. Y. Wang, N. M. Kondratiev, M. L. Gorodetsky, and T. J. Kippenberg, Nat. Photonics \textbf{8}, 145--152 (2014).
	
	\bibitem{Li2017} Q. Li, T. C. Briles, D. A. Westly, T. E. Drake, J. R. Stone, B. R. Ilic, S. A. Diddams, S. B. Papp, and K. Srinivasan, Optica \textbf{4}, 193--203 (2017).
	
	\bibitem{Suh2018} M. Suh and K. J. Vahala, Science \textbf{359}, 884--887 (2018).
	
	\bibitem{Trocha2018} P. Trocha, M. Karpov, D. Ganin, M. H. P. Pfeiffer, A. Kordts, S. Wolf, J. Krockenberger, P. Marin-Palomo, C. Weimann, S. Randel, W. Freude, T. J. Kippenberg, and C. Koos, Science \textbf{359}, 887--891 (2018).
	
	\bibitem{Marin-Palomo2017} P. Marin-Palomo, J. N. Kemal, M. Karpov, A. Kordts, J. Pfeifle, M. H. P. Pfeiffer, P. Trocha, S. Wolf, V. Brasch, M. H. Anderson, R. Rosenberger, K. Vijayan, W. Freude, T. J. Kippenberg, and C. Koos, Nature \textbf{546}, 274--279 (2017).
	
	\bibitem{Stern2018} B. Stern, X. Ji, Y. Okawachi, A. L. Gaeta, and M. Lipson, Nature \textbf{562}, 401--405 (2018).
	
	\bibitem{Vahala2004} K. Vahala, Advanced Series in Applied Physics, Vol. 5 (WORLD SCIENTIFIC, 2004) p. 839.
	
	
	\bibitem{Wang2018a} W. Wang, W. Zhang, Z. Lu, S. T. Chu, B. E. Little, Q. Yang, L. Wang, and W. Zhao, Photonics Res. \textbf{6}, 363--367 (2018).
	
	\bibitem{Wang2018b} W. Wang, Z. Lu, W. Zhang, S. T. Chu, B. E. Little, L. Wang, X. Xie, M. Liu, Q. Yang, L. Wang, J. Zhao, G. Wang, Q. Sun, Y. Liu, Y. Wang, and W. Zhao, Opt. Lett. \textbf{43}, 2002--2005 (2018).
	
	\bibitem{Herr2015} T. Herr, M. L. Gorodetsky, T. J. Kippenberg, in Nonlinear Opt. Cavity Dyn. (Wiley-VCH Verlag GmbH \& Co. KGaA, Weinheim, Germany, 2015) pp. 129--162.
	
	\bibitem{Lu2018}1. Z. Lu, W. Wang, W. Zhang, S. T. Chu, B. E. Little, M. Liu, L. Wang, C.-L. Zou, C.-H. Dong, B. Zhao, and W. Zhao, AIP Adv. \textbf{9}, 025314 (2019).
	
	\bibitem{Niu2018} R. Niu, S. Wan, S. Sun, T. Ma, H. Chen, W. Wang, Z. Lu, W. Zhang, G. Guo, C. Zou, and C. Dong, arXiv1809.06490v1 (2018).
	
	\bibitem{Guo2016} H. Guo, M. Karpov, E. Lucas, A. Kordts, M. H. P. Pfeiffer, V. Brasch, G. Lihachev, V. E. Lobanov, M. L. Gorodetsky, and T. J. Kippenberg, Nat. Phys. \textbf{13}, 94--102 (2017).
	
	\bibitem{Cao2015} Z. Cao, Z. Zhang, H.-K. Lo, and X. Ma, New J. Phys. \textbf{17}, 053014 (2015).
	
	\bibitem{Wangs2018} S. Wang, W. Chen, Z.-Q. Yin, D.-Y. He, C. Hui, P.-L. Hao, G.-J. Fan-Yuan, C. Wang, L.-J. Zhang, J. Kuang, S.-F. Liu, Z. Zhou, Y.-G. Wang, G.-C. Guo, and Z.-F. Han, Opt. Lett. \textbf{43}, 2030--2033 (2018).
	
	\bibitem{Wangxb2005} X.-B. Wang, Phys. Rev. Lett. \textbf{94}, 230503 (2005).
	
	\bibitem{Lo2005} H.-K. Lo, X. Ma, and K. Chen, Phys. Rev. Lett. \textbf{94}, 230504 (2005).
	
	\bibitem{Kues2019} M. Kues, C. Reimer, J. M. Lukens, W. J. Munro, A. M. Weiner, D. J. Moss, and R. Morandotti, Nat. Photonics \textbf{13}, 170--179 (2019).
	
\end{thebibliography}
\end{document}